\documentclass[a4paper,conference]{IEEEtran}
\usepackage{ifpdf}

\usepackage{cite}

\ifCLASSINFOpdf
\else
\fi
\usepackage{amsmath}
\usepackage{algorithmic}

\usepackage{array}
\usepackage{fixltx2e}

\usepackage{graphicx}
\usepackage{caption}
\usepackage{subcaption}
\usepackage{comment}
\usepackage{graphics,amsfonts,amsmath,amsbsy,amssymb}
\usepackage{lscape,mathptmx,helvet,makeidx,multicol,footmisc}
\usepackage{courier,type1cm,framed,subeqnarray,cite}
\usepackage{stfloats}
\usepackage{url}

\begin{document}
%
\def\x{{\mathbf x}}
\def\L{{\cal L}}

\newcommand{\lbl}[1]{\label{#1}}
\newcommand{\bftheta}{{\mbox{\boldmath$\theta$}}}
\newcommand{\bfalpha}{{\mbox{\boldmath$\alpha$}}}
\newcommand{\bflambda}{{\mbox{\boldmath$\lambda$}}}
\newcommand{\bfgamma}{{\mbox{\boldmath$\gamma$}}}
\newcommand{\bfeta}{{\mbox{\boldmath$\eta$}}}
\newcommand{\bfalphasum}{{\mbox{\boldmath$\sum\limits_{k=1}^I \alpha_k$}}}
\newcommand{\alphasum}{{\mbox{$\sum\limits_{k=1}^I \alpha_k$}}}
\newcommand{\bfTheta}{{\mbox{\boldmath$\Theta$}}}
\newcommand{\bfGamma}{{\mbox{\boldmath$\Gamma$}}}
\newcommand{\bfY}{{\mbox{\bf{Y}}}}
\newcommand{\bfD}{{\mbox{\bf{D}}}}
\newcommand{\bfa}{{\mbox{\bf{a}}}}
\newcommand{\bft}{{\mbox{\bf{t}}}}
\newcommand{\bfy}{{\mbox{\bf{y}}}}
\newcommand{\bfx}{{\mbox{\bf{x}}}}
\newcommand{\bfp}{{\mbox{\bf{p}}}}
\newcommand{\bfs}{{\mbox{\bf{s}}}}
\newcommand{\bfn}{{\mbox{\bf{n}}}}
\newcommand{\bfw}{{\mbox{\bf{w}}}}
\newcommand{\bfH}{{\mbox{\bf{H}}}}
\newcommand{\bfE}{{\mbox{\bf{E}}}}
\newcommand{\bfV}{{\mbox{\bf{V}}}}
\newcommand{\bfB}{{\mbox{\bf{B}}}}
\newcommand{\bfA}{{\mbox{\bf{A}}}}
\newcommand{\bfC}{{\mbox{\bf{C}}}}
\newcommand{\bfb}{{\mbox{\bf{b}}}}
\newcommand{\bfI}{{\mbox{\bf{I}}}}
\newcommand{\bfX}{{\mbox{\bf{X}}}}
\newcommand{\bfZ}{{\mbox{\bf{Z}}}}
\newcommand{\bfz}{{\mbox{\bf{z}}}}
\newcommand{\bfT}{{\mbox{\bf{T}}}}
\newcommand{\field}[1]{\mathbb{#1}}
\newcommand{\bfLambda}{{\mbox{\boldmath$\Lambda$}}}
\newcommand{\bfUpsilon}{{\mbox{\boldmath$\Upsilon$}}}
\newcommand{\bfOmega}{{\mbox{\boldmath$\Omega$}}}
\newcommand{\bfphi}{{\mbox{\boldmath$\phi$}}}
\newcommand{\bfPhi}{{\mbox{\boldmath$\Phi$}}}
\newcommand{\bfmu}{{\mbox{\boldmath$\mu$}}}
\newcommand{\bftau}{{\mbox{\boldmath$\tau$}}}
\newcommand{\bfsigma}{{\mbox{\boldmath$\sigma$}}}
\newcommand{\bfSigma}{{\mbox{\boldmath$\Sigma$}}}
\newcommand{\bfbeta}{{\mbox{\boldmath$\beta$}}}
\newcommand{\sumalpha}{{\mbox{\sum\limits^{I}_{k=1} \alpha_k}}}
\newcommand{\bfzero}{{\mbox{\bf{0}}}}
\newcommand{\bfone}{{\mbox{\bf{1}}}}
\newcommand{\bfPsi}{{\mbox{\boldmath$\Psi$}}}
\newcommand{\bfpi}{{\mbox{\boldmath$\pi$}}}
\newcommand{\neps}{{\mbox{$n_{\varepsilon}$}}}
\newcommand{\pval}{ \emph{p-value }\index{p-value}}
\newcommand{\pvals}{ \emph{p-values }\index{p-value}}
\newcommand{\iid}{ \emph{iid }\index{iid}}
\newcommand{\ds}{}
\newcommand{\singspace}{\renewcommand{\baselinestretch}{1}\small\normalsize}
\newcommand{\doubspace}{\renewcommand{\baselinestretch}{2}\small\normalsize}
\newcommand{\methodhead}[1]{\pagebreak[3]\vspace*{0.3in}\noindent{\em{\underline{#1}}} \nopagebreak[5]\newline \nopagebreak[5]~

}
\newcounter{partcount}
\newcommand{\myhyp}[2]{
~\\
\hspace*{0.5in} \ensuremath{\mathbf{#1}}\nopagebreak[4]\\
\hspace*{0.5in} \ensuremath{\mathbf{#2}}\\\nopagebreak[4]
~\\ }
\newcommand{\mypart}[1]{
   \addtocounter{partcount}{1}
    \cleardoublepage
    \vspace*{2in}
    \noindent
    {\huge {\bf Part \Roman{partcount}
    ~\\~\\~\\~\\}}
    {\Huge {\bf #1}\\}
    \addcontentsline{toc}{part}{Part \Roman{partcount}: ~#1}
    \vspace*{1in}\\

        }
\newcounter{chapcount}
\newcommand{\alphachapter}[1]{
   \stepcounter{chapcount}
\addtocounter{chapter}{0}
    \setcounter{section}{0}

    \cleardoublepage
    \vspace*{2in}
    \noindent
    {\huge {\bf Chapter \Alph{chapcount}
    ~\\~\\~\\~\\}}
    {\Huge {\bf #1}\\}
    \addcontentsline{toc}{chapter}{\Alph{chapcount} ~#1}
    \vspace*{1in}\\
        }
\newcounter{mychapcounter}
\newcommand{\mychap}[1]
{
    \stepcounter{mychapcounter}
    \refstepcounter{chapter}
    \cleardoublepage
    \vspace*{2in}
    \noindent
    {\huge {\bf Chapter \arabic{mychapcounter}~\\
    ~\\~\\~\\}}
    {\Huge {\bf #1}\\}
    \addcontentsline{toc}{chapter}{\arabic{mychapcounter} ~#1}
    \vspace*{1in}\\
}
\title{Statistical Methods for Assessing Differences in False Non-Match Rates Across Demographic Groups}

\author{\IEEEauthorblockN{Michael Schuckers}
\IEEEauthorblockA{Mathematics, Computer Science and Statistics\\
St.~Lawrence University\\
Canton, NY, USA\\
Email: \texttt{schuckers@stlawu.edu}}
\and
\IEEEauthorblockN{Sandip Purnapatra, Kaniz Fatima, Daqing Hou, Stephanie Schuckers}
\IEEEauthorblockA{Computer and Electrical Engineering, \\Clarkson University\\
Potsdam, NY, USA\\
Email: \texttt{\{purnaps, fatimak, dhou, sschucke\}@clarkson.edu}}
}


%


\maketitle

\begin{abstract}
Biometric recognition is used across a variety of applications from cyber security to border security. Recent research has focused on ensuring biometric performance (false negatives and false positives) is fair across demographic groups. While there has been significant progress on the development of metrics, the evaluation of the performance across groups, and the  mitigation of any problems, there has been little work incorporating statistical variation.  This is important because differences among groups can be found by chance when no difference is present.  In statistics this is called a Type I error.  
Differences among groups may be due to sampling variation or they may be due to actual difference in system performance.  Discriminating between these two sources of error is essential for good decision making about fairness and equity.  This paper presents two novel 
statistical approaches for assessing fairness across demographic groups.  The first methodology is a bootstrapped-based hypothesis test, while the second is simpler test methodology focused upon non-statistical audience. For the latter we 
present the results of a simulation study about the relationship between the margin of error and factors such as number of subjects, number of attempts, correlation between attempts, underlying false non-match rates(FNMR's), and number of groups.\\
\end{abstract}


%
\IEEEpeerreviewmaketitle

\section{Introduction}

Biometric recognition is a technology that has broad application for border security, e-commerce, financial transactions, health care, and benefit distribution.  With its explosion in use, there are concerns about the fairness of solutions across the broad spectrum of individuals, based on factors such as age, race, ethnicity, gender, education, socioeconomic status, etc.  In particular, since biometric recognition has a possibility of error, both false negatives (false rejection) and false positives (false acceptance), the expectation is that solutions have performance which are ``fair'' across demographic groups. Buolamwini, et. al. found that gender classification based on a single face image had a higher error rate for darker-skinned females with a high 34.7\% error rate, compared to other  groups (intersections of skin types and genders) \cite{gender-shades-1}, \cite{gender-shades-2}.  While focused on gender classification rather then face recognition,  these papers brought considerable attention to this issue. 
 Others found demographic differences in face recognition for some algorithms and systems \cite{FRVT-demographics}, \cite{Demographic-Effects}.   

To quantify the equitability of the various face recognition algorithms, multiple metrics have been proposed to evaluate fairness. ProposedFairness Discrepancy Rate (FDR) weights the two types of errors seen in biometric recognition (false accept and false reject rates), either equally or otherwise, and balances FDR across groups \cite{FDR}. The U.S.~National Institute of Standards and Technology (NIST) introduced the Inequity Rate (IR) metrics for face recognition algorithm performance testing and \cite{howard2022evaluating} proposed two interpretibility criterion for biometric systems, i.e.~Functional Fairness Measure Criteria (FFMC) and Gini Aggregation Rate for Biometric Equitability (GARBE).     In other artificial intelligence (AI) research,  
  evaluation metrics include demographic parity, equalized odds, and equal opportunity \cite{AI-machine-learning} \cite{ibm-toolkit} \cite{AI-fairness360} \cite{Algorithmic-Equity}.  

However, with all of these active research and analyses, there has been limited contribution towards recommending appropriate statistical methods for determining when two or more groups are ``equal'' or not.  This is essential, as any metric when measured in a sample, will have uncertainty which is a function of variability, correlation, number of groups, and other factors.  This uncertainty can be measured through statistical methods, e.g.~ confidence intervals, to determine the likelihood that differences are found by chance or are a true difference.  Given that exact ``equality'' is unlikely, if not impossible, for a set of groups, these methods allow for appropriate conclusions to be drawn from results.

This paper focuses on statistical methods for fairness solely for false negatives.  Biometric solutions used widely by the public are typically based on ``verification'' or one-to-one matching.  A false negative error is when the correct individual is falsely rejected, e.g., does not match their enrollment on a mobile device, passport, bank, or government benefits provider. This ``error'' may block an individual from accessing benefits which they are entitled to. The goal of this paper is to consider the statistical methods to address differences in false non-match rates based on number of subjects, number of attempts, correlation in attempts, and number of possible demographic groups in the test.  The number of subjects and number of attempts can decrease the variability as the number of subjects and attempts increase; whereas, incorporating the correlation between number of attempts may increase sample variability.  Most importantly, the number of demographic groups being compared impacts the variation as an increased number of groups increases the chances that a difference between groups may be found ``by chance'', and thus adjustments need to be made in the test due to this effect, often called multiplicity \cite{multcomp}. 

This paper develops two approaches to detecting differences in FNMR's between demographic groups.  Additionally, we explore the trade-off among variation parameters based on simulations of a hypothetical equity study.  In addition to giving guidance on expected outcomes of such a study, this paper will provide suggestions for ``practical'' thresholds that could be used for when to say that a group is different that would minimize the possibility that that difference was based upon chance alone.    In the next
section we discuss related statistical work that has been done to assess differences in FNMR's between groups.  
Section III introduces the basic statistical structures needed to estimate variation in FNMR estimation.  A bootstrap hypothesis test for the equality of FNMR across $G$ groups is presented in Section IV, as well as a simplified alternative that yields a margin of error for detecting differences among groups.  That section also includes results of a simulation study.  We summarize and discuss this work and possible alternatives in Section V.

\section{Related Work}


In this section, we discuss other work on statistical methods for comparison of bioauthentication across demographic groups.  The NIST Information Technology Laboratory (ITL) quantifies the accuracy of face recognition algorithms for the demographic groups of sex, age, and race~\cite{FRVT-demographics}. A component of the evaluation focuses on FNMR for one-to-one verification algorithms on four large datasets of photographs collected in U.S. governmental applications (domestic mugshots, immigration application photos, border crossing, and visa applications). 
For high-quality photos, FNMR was found to be low and it is fairly difficult to measure false negative differentials across demographics. Compared to high-quality application photos, the FNMR is higher for lower-quality border crossing images. 
Similar observations regarding image quality have been made by others, e.g. \cite{Demographic-Effects}. 
A measure of uncertainty is calculated for each demographic group based on a bootstrapping approach.  In bootstrapping, the genuine scores are sampled 2000 times and the 95\% interval is plotted providing bounds for each group.  No method was presented to suggest when an algorithm might be "fair" under uncertainty.  A notional approach might be to declare an algorithm fair if the intervals plotted overlap across all combinations of groups.  This, however, does not fully address the possibility of Type I errors.

Cook et al. \cite{Demographic-Effects} examined the effect of demographic factors on the performance of the eleven commercial face biometric systems tested
as part of the 2018 United States Department of Homeland Security, Science and Technology Directorate (DHS S\&T) Biometric Technology Rally. Each participating system  was tasked with acquiring face images from a diverse population of 363 subjects in a controlled environment. Biometric performance was assessed by measuring both efficiency (transaction times) and accuracy (mated similarity scores using a leading commercial algorithm). 
The authors quantified the effect of relative facial skin reflectance and other demographic covariates on performance using linear modeling.
Both the efficiency and accuracy of the tested acquisition systems were significantly affected by
multiple demographic covariates including skin reflectance, gender, age, eyewear, and height, with skin reflectance having the strongest net
linear effect on performance. 
Linear modeling showed that lower (darker) skin reflectance was associated with lower efficiency (higher transaction times) and accuracy (lower mated similarity scores) \cite{Demographic-Effects}. 
While statistical significance of demographic factors was considered based on a linear model of match scores, this approach may not be applicable for assessing commercial systems which operate at a fixed threshold.


de Freitas Pereira and Marcel \cite{FDR} introduce the Fairness Discrepancy Rate (FDR) which is a summary of system performance accounting for both FNMR and FMR.  
Their approach uses a ``relaxation constant'' rather than trying to assess the sampling variation or statistical variation between FNMR's from different demographic groups.
Howard et al. \cite{howard2022evaluating} present an evaluation of FDR noting its scaling problem. To address this scaling problem, the authors propose a new fairness measure called Gini Aggregation Rate for Biometric Equitability (GARBE).

Other research has also performed extensive evaluations of face recognition across demographic groups, e.g. \cite{skin-tone}, but have not presented statistical methods as part of their work.

\section{Variance and Correlation Structure of FNMR}
Statistical methods for estimation of FNMR's are dependent upon the variance and correlation of matching decisions.
In this section, we present the basic statistical structures for a single FNMR following \cite{schuckers10}.  This structure forms the basis for the statistical methods that we present in the next sections.  
Let $D_{iij}$ represent the decision for the $j^{th}$ pair of captures
or signals collected on the $i^{th}$ individual, where $n$ is the
number of individuals, $i=1,\ldots,n$ and $j=1,\ldots,m_i$. Thus,
the number of sample pairs that are compared for the $i^{th}$
individual is $m_i$, and $n$ is the number of different individuals
being compared. The use of $m_i$ implies that we are allowing
the number of comparisons made per individual to vary across
individuals.  We then define
\begin{equation}
D_{iij} = \left\{ \begin{array}{rl}
  1  & \mbox{if $j^{th}$ pair of signals from individual $i$} \\
     &\mbox{is declared a non-match},\\
  0 & \mbox{otherwise}.\\
\end{array}
\right.
\end{equation}
We assume for the $D_{iij}$'s that
$E[D_{iij}]=\pi$ and $V[D_{iij}]=\pi(1-\pi)$
 represent
the mean and variance, respectively. Thus, $\pi$ \index{$\pi$@pi}\index{pi@$\pi$}represents the FNMR.  We assume
that we have a stationary \index{stationary} matching process
within each demographic group and implicit in this assumption is that we have a fixed threshold within each group.  
Our estimate of $\pi$, the process FNMR, will
be the total number of errors divided by the total number of
decisions:
\begin{equation}\lbl{fnmr:pihat}
\hat{\pi}=[\sum_{i=1}^{n} \sum_{j=1}^{m_i} D_{iij}]/
[\sum_{i=1}^{n} m_i].
\end{equation}

 Following  Schuckers \cite{schuckers08, schuckerstifs}, we have
 the following correlation structure for the $D_{iij}'s$:
\begin{equation}
\lbl{corr.fnmr} Corr(D_{iij},D_{i'i'j'})=\left\{
\begin{array}{ccl}
  1 &\mbox{if}& i=i',j=j' \\
  \rho &\mbox{if}& i=i', j \ne j' \\
  0&&otherwise.
\end{array}
\right.
\end{equation}
This correlation structure for the FNMR is based
upon the idea that the there will only be correlations between
decisions made on signals from the same individual but not between
decisions made on signals from different individuals.
  Thus, conditional upon the error rate, there is no
correlation between decisions on the $i^{th}$ individual and
decisions on the $i'^{ ~ th}$ individual, when $i \neq i'$.
The degree of correlation is summarized by $\rho$.

Then we can write the variance of $\hat{\pi}$, the estimated
FNMR, as
\begin{equation}
\lbl{VFNMR} V[\hat{\pi}]= N_{\pi}^{-2}\pi(1-\pi)[N_{\pi} + \rho
\sum_{i=1}^{n} m_i(m_i-1)]
\end{equation}
where $N_{\pi} = \sum_{i=1}^{n} m_i$.  An estimator for $\rho$ is given by:
\begin{equation}
\lbl{corr.fnmr.estimate}
\hat{\rho}=\frac{\ds \sum_{i=1}^{n} \sum_{j=1}^{m_i} \sum_{j'=1 \atop j'\neq j}^{m_i}
(D_{iij}-\hat{\pi})(D_{iij'}-\hat{\pi})}{\ds \hat{\pi}(1-\hat{\pi})\sum_{i=1}^{n} m_i(m_i-1)}
.
\end{equation}

Models like that found in (\ref{corr.fnmr}) are known as intra-individual or intra-class
models and have been studied extensively in the statistics
literature, e.g. Fleiss et al. \cite{fleiss03}, Williams
\cite{williams75} or Ridout et al. \cite{ridout99}. The
parameter $\rho$ in the models above represents the intra-class
correlation. This measures the degree of similarity between
the decisions made on each individual. If the decisions on each
individual are varying in a way that suggests that the decisions are
not dependent upon the individual then $\rho$ is zero, meaning that
the observations are uncorrelated. Negative values of $\rho$ are
possible but such values suggest that decisions on signals from the
same individual are less similar to each other than they are to all
of the other decisions. This seems unlikely to be the case in the
context of biometric authentication. Several authors, including
Fleiss et al. \cite{fleiss03}, have suggested using the following
alternative way of writing (\ref{VFNMR})
\begin{equation}
\lbl{var.fmr.approx} V[\hat{\pi}]=
N_{\pi}^{-1}\pi(1-\pi)(1+(m_0-1)\rho)
\end{equation}
where $
m_0=\frac{ \sum_{i=1}^{n} m_i^2}{ N_\pi}.$
  If $m_i=m$ for all $i$, then $N_{\pi}=nm$ and the variance of
$\hat{\pi}$ from (\ref{var.fmr.approx}) becomes $
\lbl{approx.vfnmr}
V[\hat{\pi}]= (nm)^{-1}\pi(1-\pi)(1+(m-1)\rho).$

The intra-class correlation has a direct relationship with the variance of
$\hat{\pi}$.  As $\rho$
increases, the variance in both cases increases.  This is a
consequence of the lack of independent information from each
individual.  If $\rho$ is large, then  each additional decision on a
previously observed individual is providing little new information.

\section{Statistical Methods for Multiple FNMR's}
To evaluate and assess if different FNMR's are \textit{detectably} different\footnote{We are using detectably different here in place of significantly different. See \cite{asa_pvalues}.}, we need to understand the variation due to sampling.  In equity studies across different demographic groups, we need to account for the sampling variation from each of the $G$ groups.  For what follows we will assume that there are $G$ demographic groups across multiple dimensions.  For example, if a study wants to compare four ethnic groups, five education levels, three genders and five age groups, then $G=4+5+3+5=17$.
Methods for comparisons of different demographic groups on their FNMR's generally involve comparing FNMR's across three or more categories.  These methods are more advanced and more sophisticated than those for comparing two groups or for comparing a single group to a specific value.  
See \cite{schuckers10} for methods involving one or two FNMR's.  Below we present and discuss statistical methods for determining if there are detectable differences between FNMR's among $G$ independent groups. This single methodology is preferable to testing multiple times which yields potentially higher rates of Type I errors.  
Below we begin with a 
bootstrap hypothesis test and that is followed by a simplified version that may be
more easily understood by a broad audience.  

\subsection{Bootstrap Hypothesis Test}
Since the individuals and decisions are independent between
groups, we bootstrap each group separately to mirror the variability in the sampling process.  
As with an analysis of variance (ANOVA), we use a test statistic similar to
the usual F-statistic and then we compare the observed value to a
reference distribution composed of bootstrapped values.  Formally, our hypotheses are: $H_0:\pi_1=\pi_2=\pi_3=\ldots=
\pi_G,\mbox{ vs } H_1: \mbox{ at least one }\pi_g \mbox{ is different }$.

\begin{enumerate}
\item Calculate
\begin{equation}
\lbl{fnmr.fstat}
F=\frac {\left[ \sum_{g=1}^{G}  N_{\pi}^{(g)}
(\hat{\pi}_g- \hat{\pi})^2\right]/(G-1)}{\left[
\sum_{g=1}^{G} N_{\pi}^{(g)}
\hat{\pi}_g(1-\hat{\pi}_g)(1+(m_0^{(g)}-1)\hat{\rho}_g)
\right]/(N-G)}
\end{equation}
 for the observed data where
 \begin{equation}
 \lbl{fnmr.avg}
 \hat{\pi}=\frac{\ds
\sum_{g=1}^{G}  N^{(g)}_{\pi} \hat{\pi}_g}{\ds
\sum_{g=1}^{G}  N^{(g)}_{\pi}}, \mbox{ }
 \hat{\pi}_g=\frac{ \sum_{i=1}^{n_g} \sum_{j=1}^{m^{(g)}_i} D^{(g)}_{i i j}}
 { \sum_{i=1}^{n_g} m^{(g)}_i}
\end{equation}
and $N=\ds \sum_{g=1}^{G} N_{\pi}^{(g)}$.  Here $\hat{\pi}$ is the (weighted) average
of the FNMR's across the $G$ groups. 

 \item For each group $g$, sample $n_g$ individuals with replacement from the $n_g$ individuals
in the $g^{th}$ group.  Denote these selected individuals
by $b^{(g)}_1,b^{(g)}_2,\ldots,b^{(g)}_{n_g}$.
For each selected individual, $b^{(g)}_i$, in the $g^{th}$ group
take all the $m_{b^{(g)}_i}$ non-match decisions for that individual.
 Call these selected decisions $D^{(g)}_{b^{(g)}_i b^{(g)}_i j}$'s with $j=1
 ,\ldots, m_{b^{(g)}_i}$ and calculate
 \begin{equation}
 \label{eq:bootstrapdiff}
 \hat{\pi}_g^{b}=\frac{  \sum_{i=1}^{n_g}
 \sum_{j=1}^{m_{b^{(g)}_i}} D^{(g)}_{b^{(g)}_i b^{(g)}_i j}}
 { \sum_{i=1}^{n_g} m_{b^{(g)}_i}} - \hat{\pi}_g + \hat{\pi}.
 \end{equation}

\item Repeat the previous two steps some large number of times, $K$, each time
calculating and storing
\begin{equation}
F_{\pi}=\frac {\left[ \sum_{g=1}^{G}  N_{\pi}^{(g)}
(\hat{\pi}^b_g- \bar{\pi}^b)^2\right]/(G-1)}{\left[
\sum_{g=1}^{G} N_{\pi}^{(g)}
\hat{\pi}^b_g(1-\hat{\pi}^b_g)(1+(m_0^{(g)b}-1)\hat{\rho}^b_g)
\right]/(N-G)}.
\end{equation}
Here $\bar{\pi}^b$ represents the calculations given
above applied to the bootstrapped matching decisions,
 \begin{equation}
 \bar{\pi}^b=\frac{
\sum_{g=1}^{G}  N^{(g)b}_{\pi} \hat{\pi}^b_g}{
 \sum_{g=1}^{G}  N^{(g)b}_{\pi}},
\end{equation}
where $\ds N^{(g)b}_{\pi}=\sum_{i=1}^{n_g} m_{b^{(g)}_i}$.  The values for
$\hat{\rho}^b_g$ and $m_0^{(g)b}$ are found by using the usual estimates
for those quantities applied to the bootstrapped decisions from the $g^{th}$
group.

\item Then the \pval for this test is
$
p=\frac{\ds 1+\sum_{\varsigma=1}^{K}I_{\{F_{\pi} \geq F\}}}{K+1}.
$

\item We will conclude that at least one of the FNMR's is different from the rest if the \pval is small.  When
a significance \index{significance level}level is designated, then we
will reject the null hypothesis, $H_0$, if $p <\alpha$.

\end{enumerate}
We adjust our bootstrapped
sample statistic, here $\hat{\pi}^b$, to center their distributions of the 
FNMR's in each group in accordance
with the null hypothesis of equality between the $G$ FNMR's.
In this case we center with respect to our estimate of the
FNMR, $\hat{\pi}$, assuming all of the FNMR's are identical.

\begin{figure}
    \centering
    \includegraphics[width=.4\textwidth]{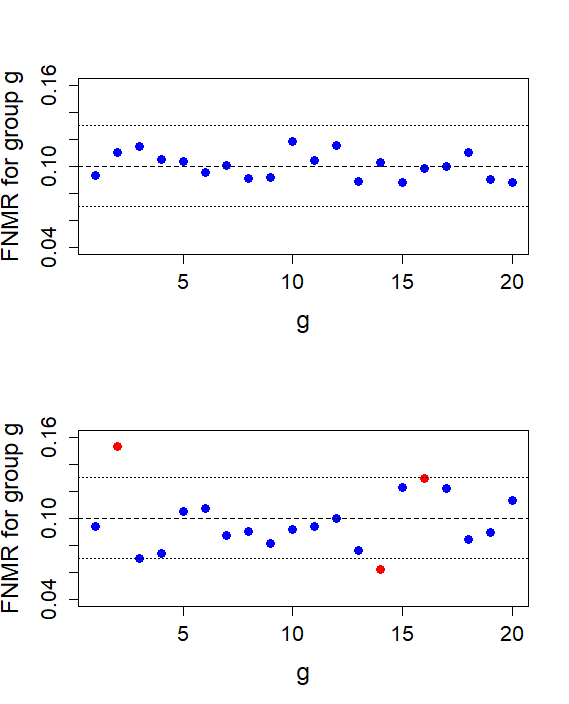}
    \caption{The top subfigure has all $G=20$ group FNMR's fall between the bounds (dotted lines) generated by 
    adding and subtracting a margin of error, $M$, from the overall FNMR (dashed lines), while the bottom subfigure has three subgroups (in red) that fall outside of these bounds.}
    \label{fig:illustrateM}
\end{figure}

\subsection{Simplified Alternative Methodology for Broad Audience Reporting}
The methods of the previous subsection may be difficult to explain to a broad, non-technical audience.  Consequently, in this section, we propose a methodology for simplifying the testing of multiple FNMR's across demographic groups. That is, we will conclude that a particular subgroup $g$ has a different FNMR if its observed FNMR is outside of the interval created by taking the average FNMR, $\hat{\pi}$, and adding and subtracting a margin of error, $M$.  
This methodology is more straightforward for explaining to decision makers and to wide audiences and
takes advantage of the common usage of the margin of error.  
\begin{figure*}[!t]
\centering
\textbf{Simulation Study Results for Margin of Error (M) versus Number of Groups (G)}

\begin{minipage}[htbp]{0.3\textwidth}
    \includegraphics[width=\textwidth]{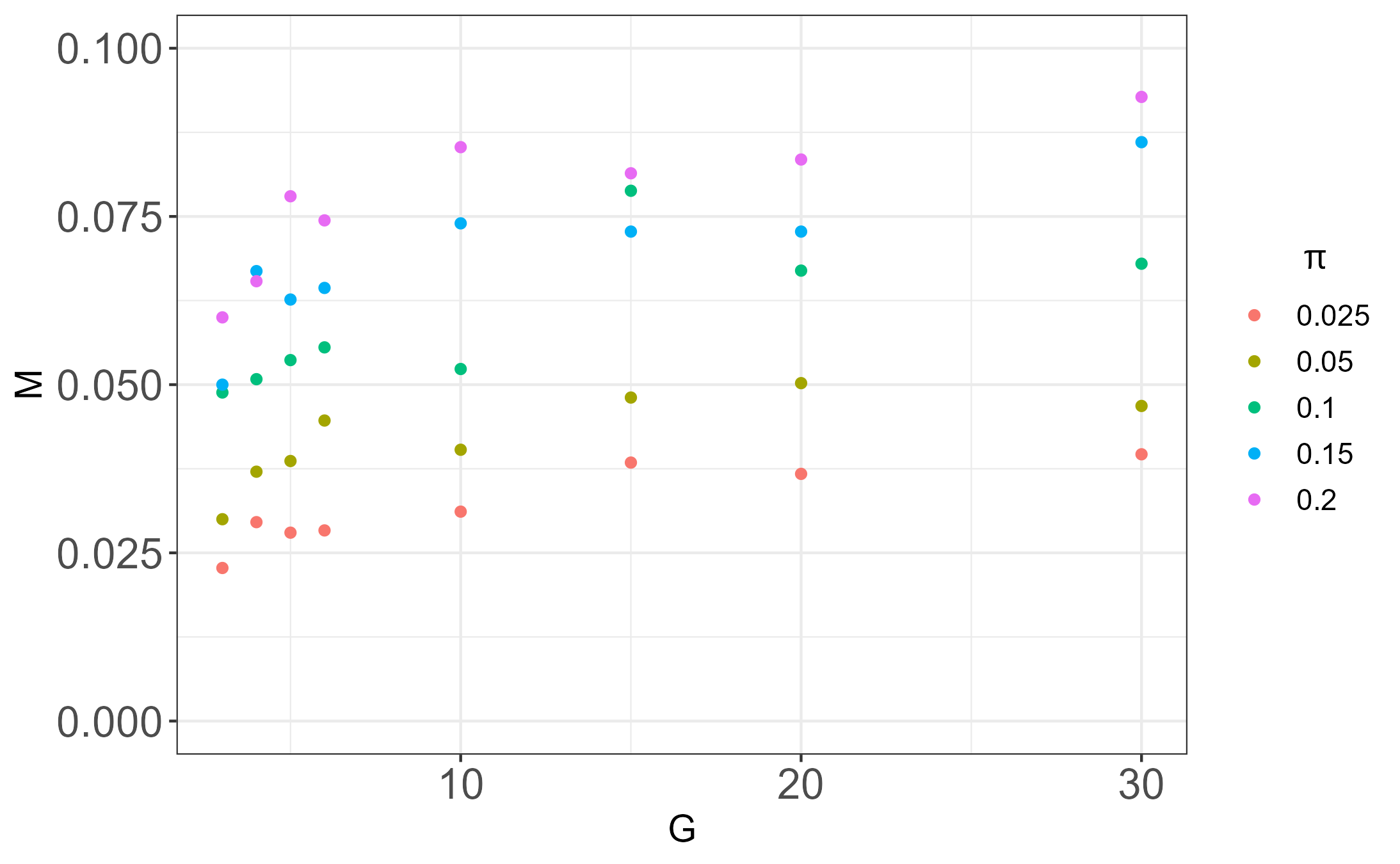}
       \hspace*{.4in}
       {\small (a) n=100, m=2, $\rho$=0.05}
    \label{S2_Corr0.05_100}
    \end{minipage}
 \begin{minipage}[htbp]{0.3\textwidth}
    \includegraphics[width=\textwidth]{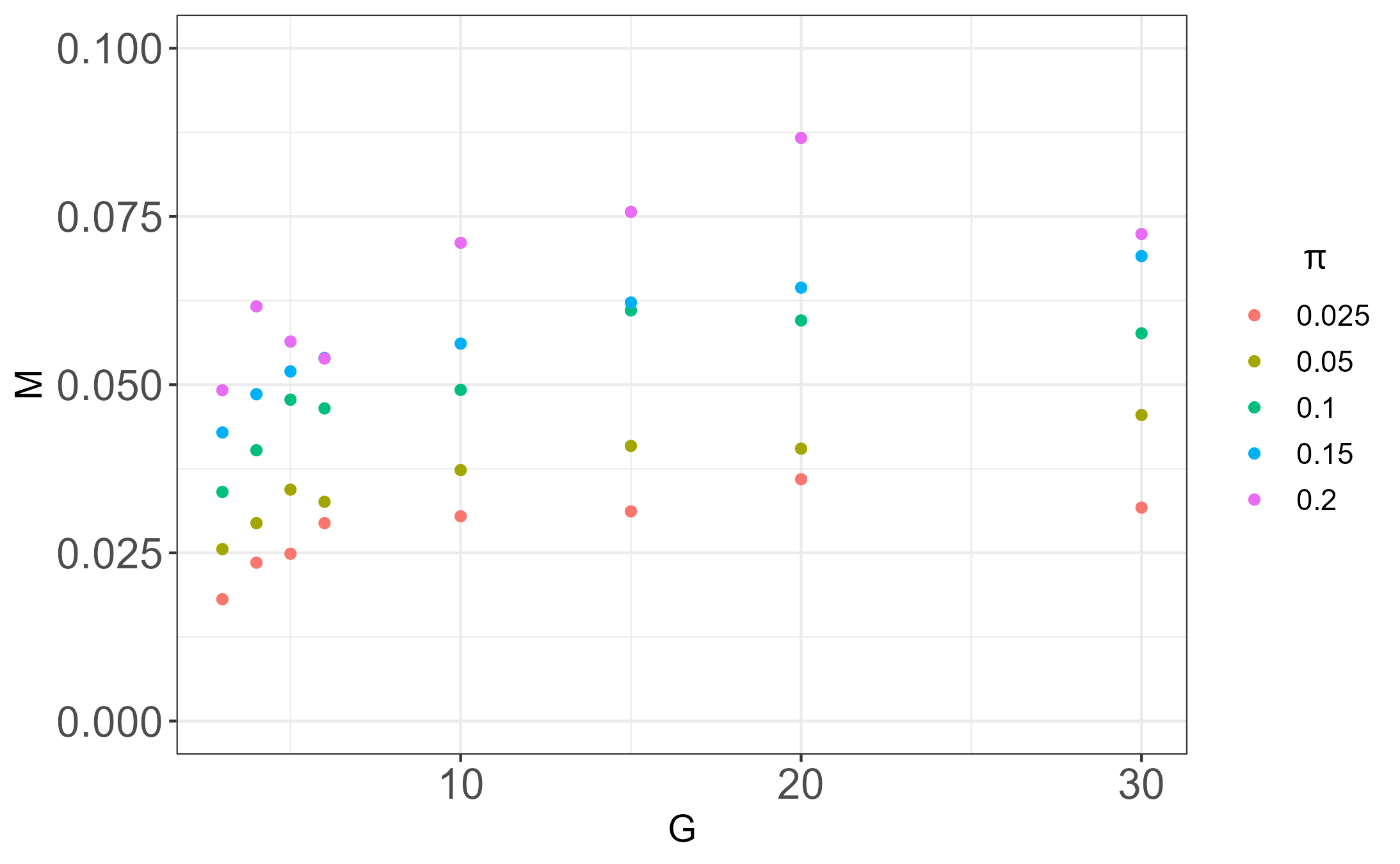}
      \hspace*{.4in}{\small (b) n=100, m=3, $\rho$=0.05}
    \label{fig:S3_Corr0.05_100}
    \end{minipage}
\begin{minipage}[htbp]{0.3\textwidth}
    \includegraphics[width=\textwidth]{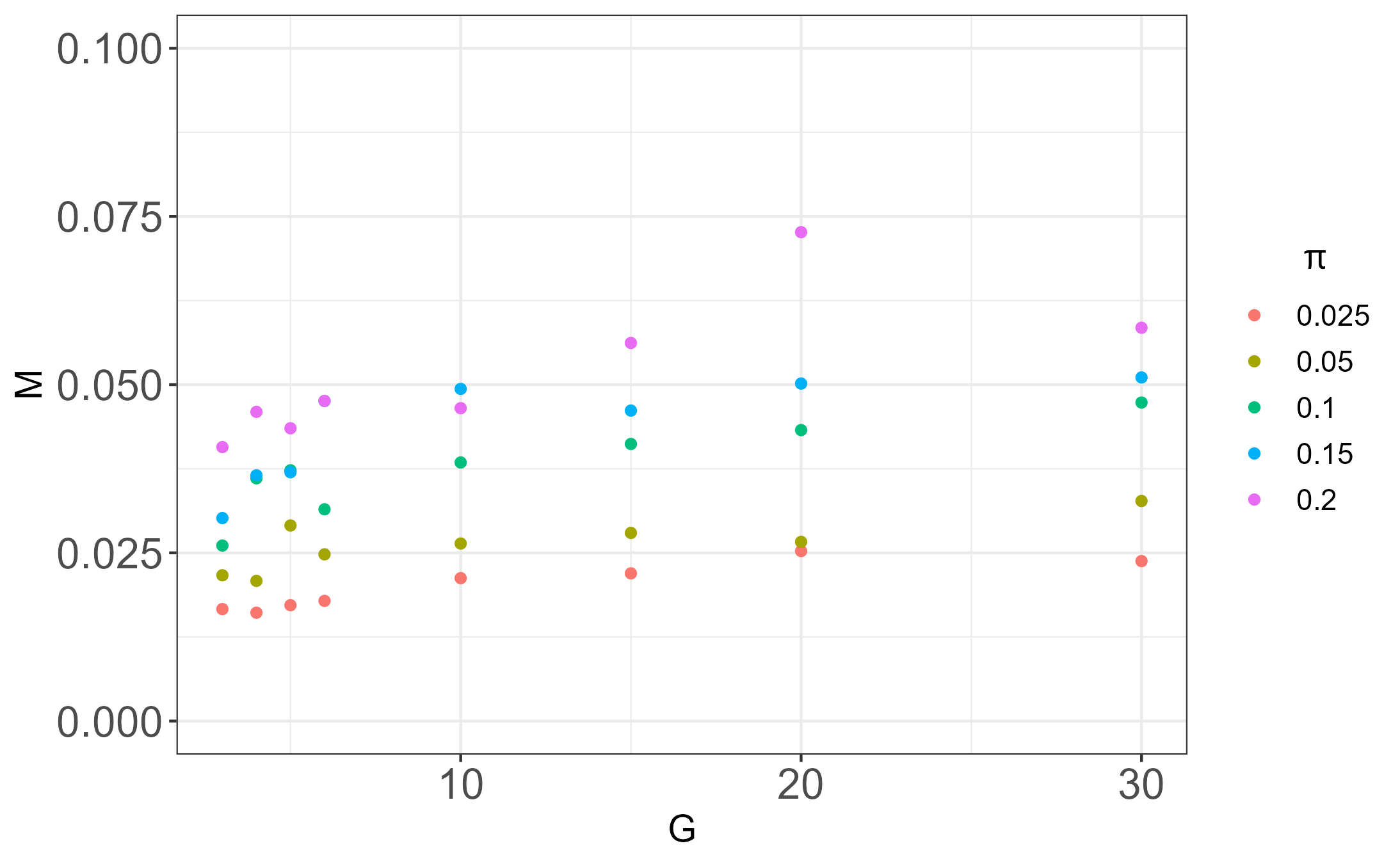}
          \hspace*{.4in} {\small (c) n=100, m=6, $\rho$=0.05}
    \label{fig:S6_Corr0.05_100}
\end{minipage}
 \hfill
 
\vspace{0.2in}
\begin{minipage}[htbp]{0.3\textwidth}
    \includegraphics[width=\textwidth]{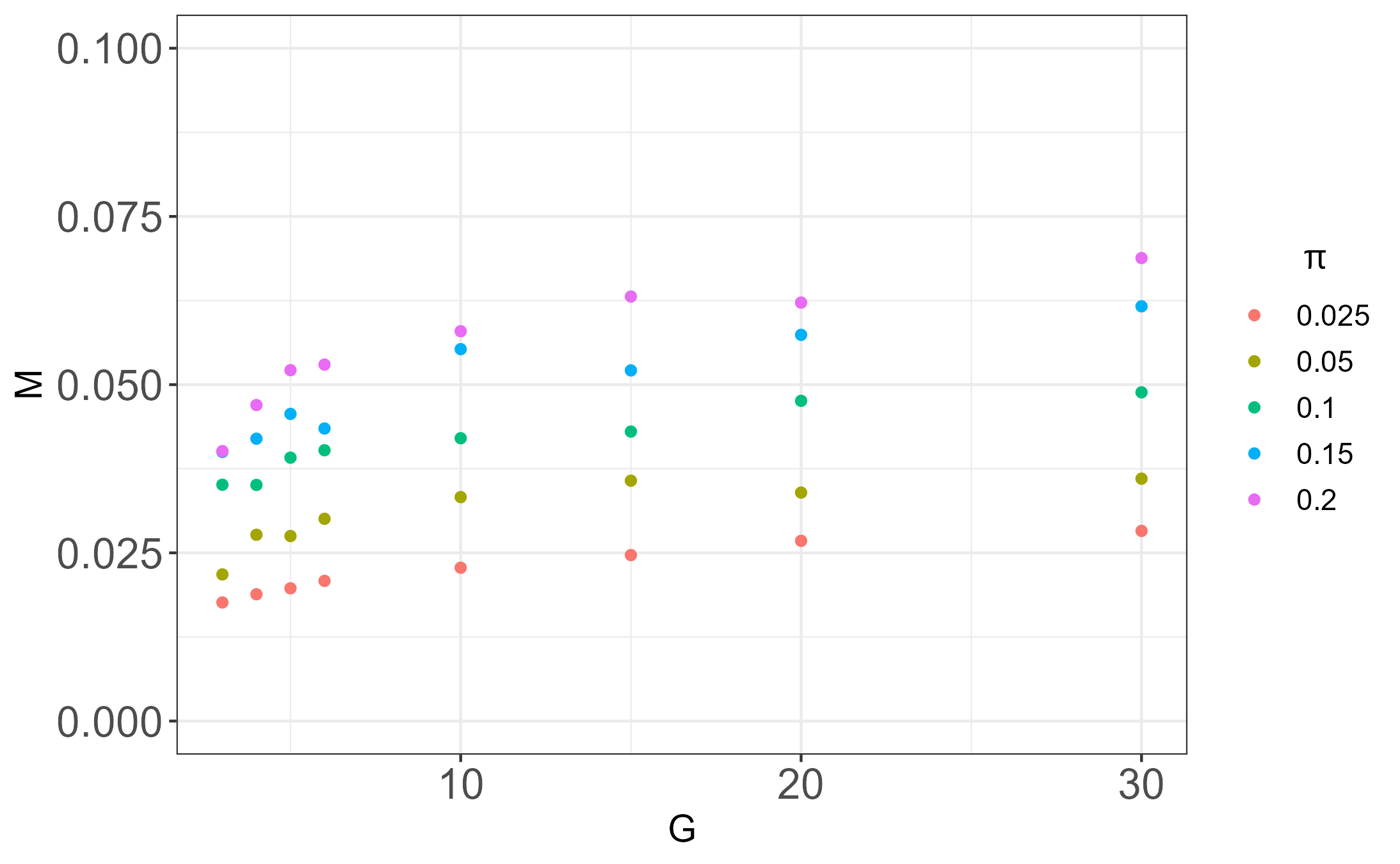}
          \hspace*{.4in}{\small (d) n=200, m=2, $\rho$=0.05}
    \label{fig:S2_Corr0.05_200}
\end{minipage}
 \begin{minipage}[htbp]{0.3\textwidth}
    \includegraphics[width=\textwidth]{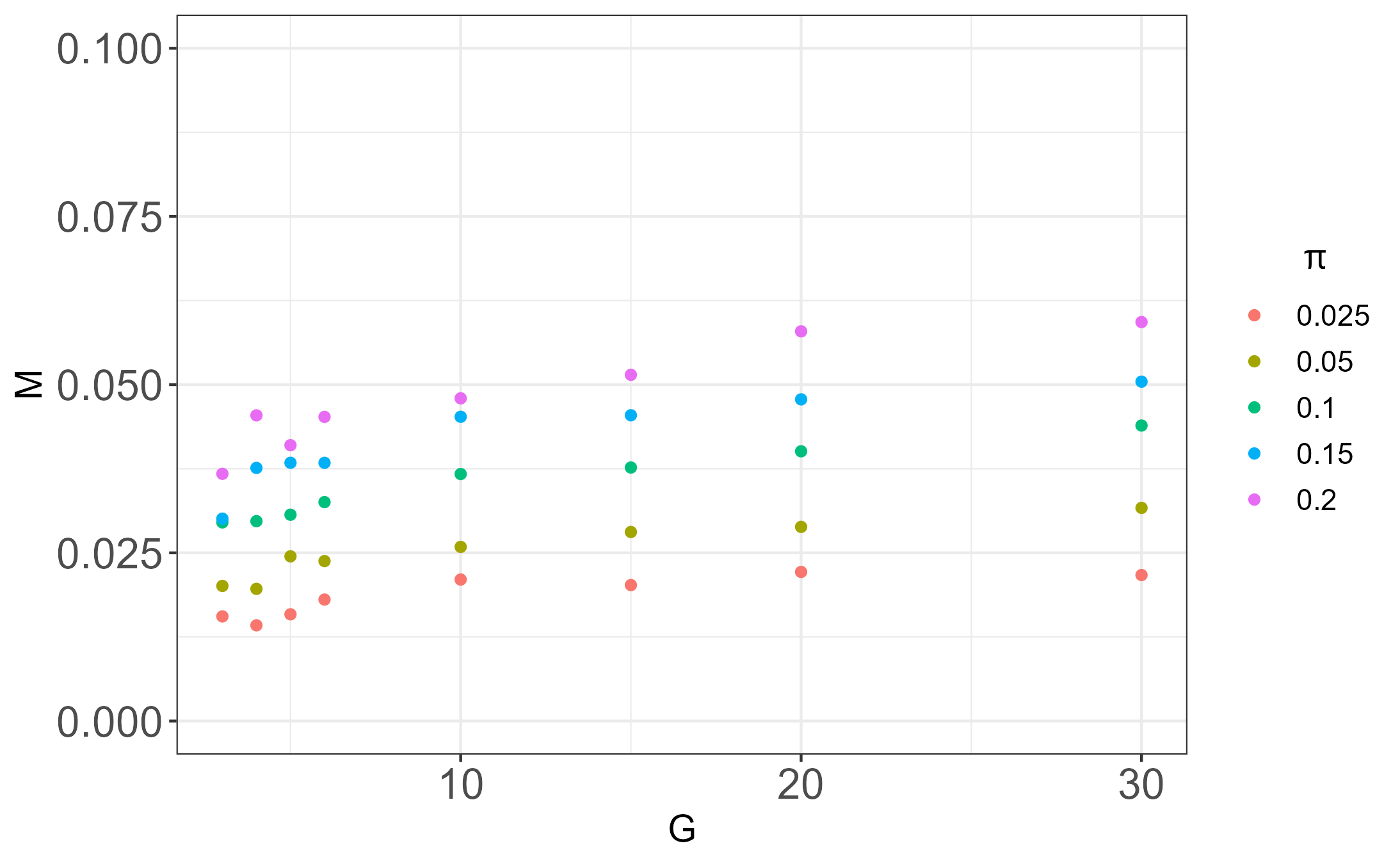}
         \hspace*{.4in}{\small (e) n=200, m=3, $\rho$=0.05}
    \label{fig:S3_Corr0.05_200}
    \end{minipage}
\begin{minipage}[htbp]{0.3\textwidth}
    \includegraphics[width=\textwidth]{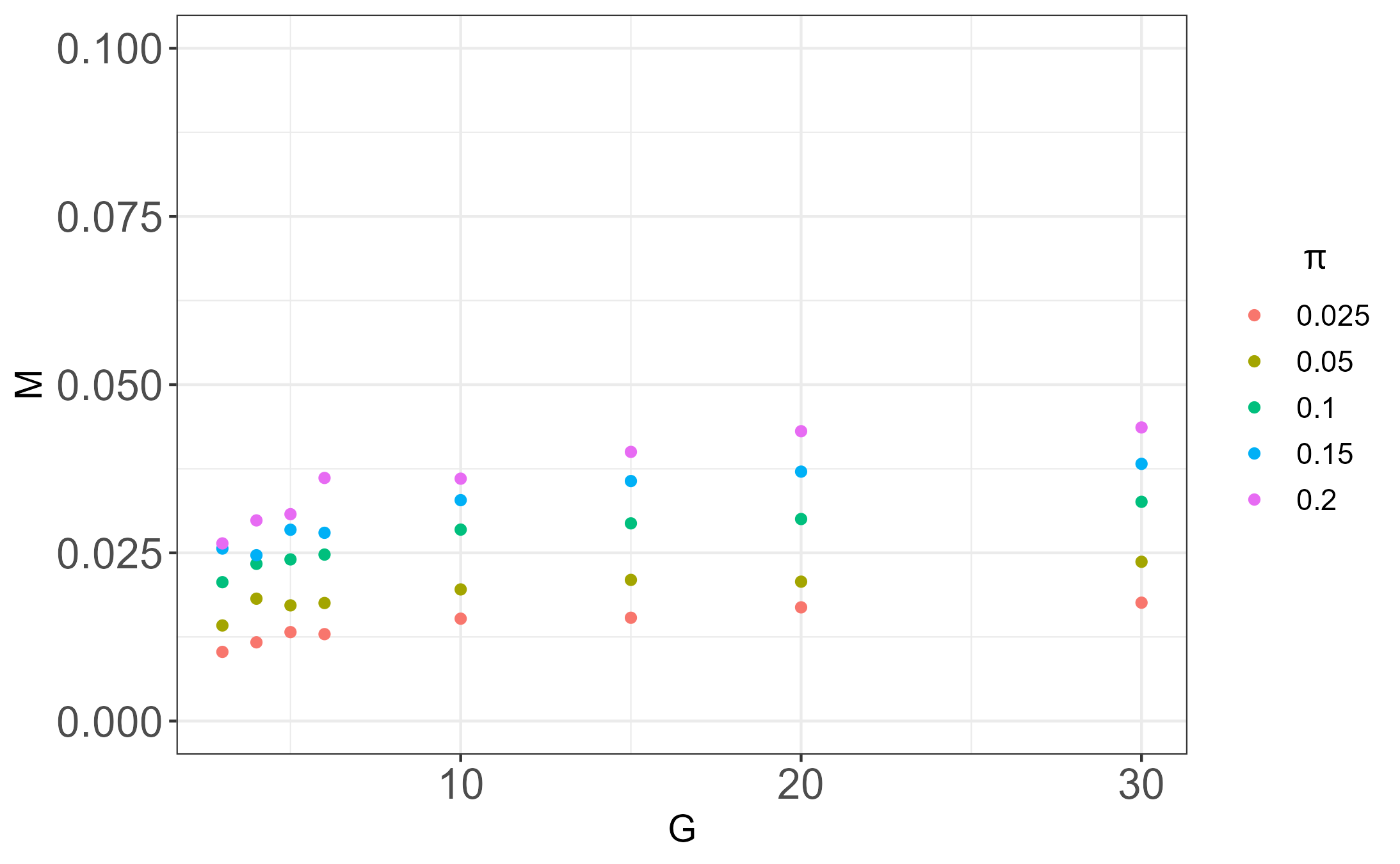}
          \hspace*{.4in}{\small (f) n=200, m=6, $\rho$=0.05}
    \label{fig:S6_Corr0.05_200}
\end{minipage}
\hfill

\vspace{0.2in}
\begin{minipage}[htbp]{0.3\textwidth}
    \includegraphics[width=\textwidth]{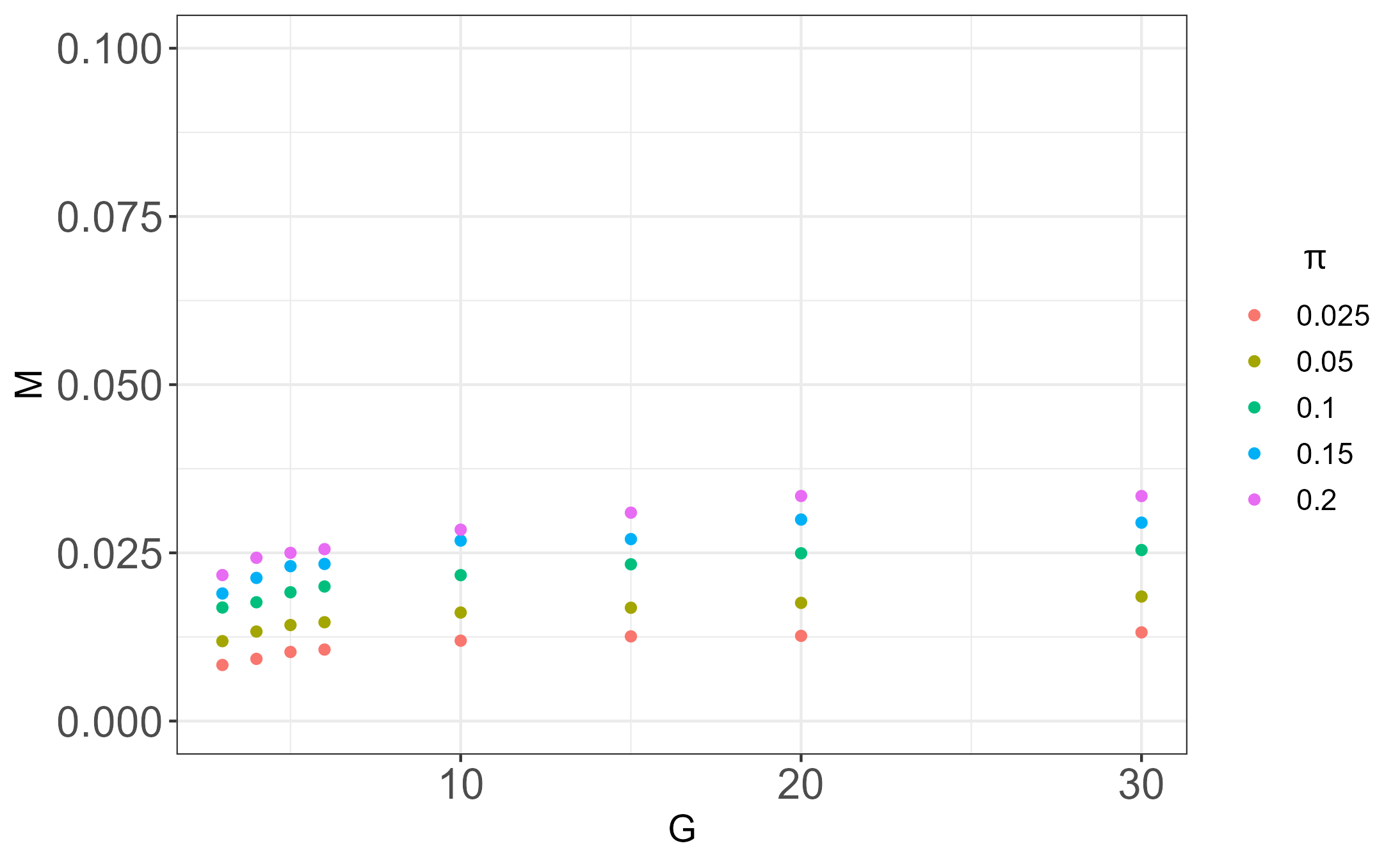}
          \hspace*{.4in}{\small (g) n=800, m=2, $\rho$=0.05}
    \label{S2_Corr0.05_800}
\end{minipage}
 \begin{minipage}[htbp]{0.3\textwidth}
    \includegraphics[width=\textwidth]{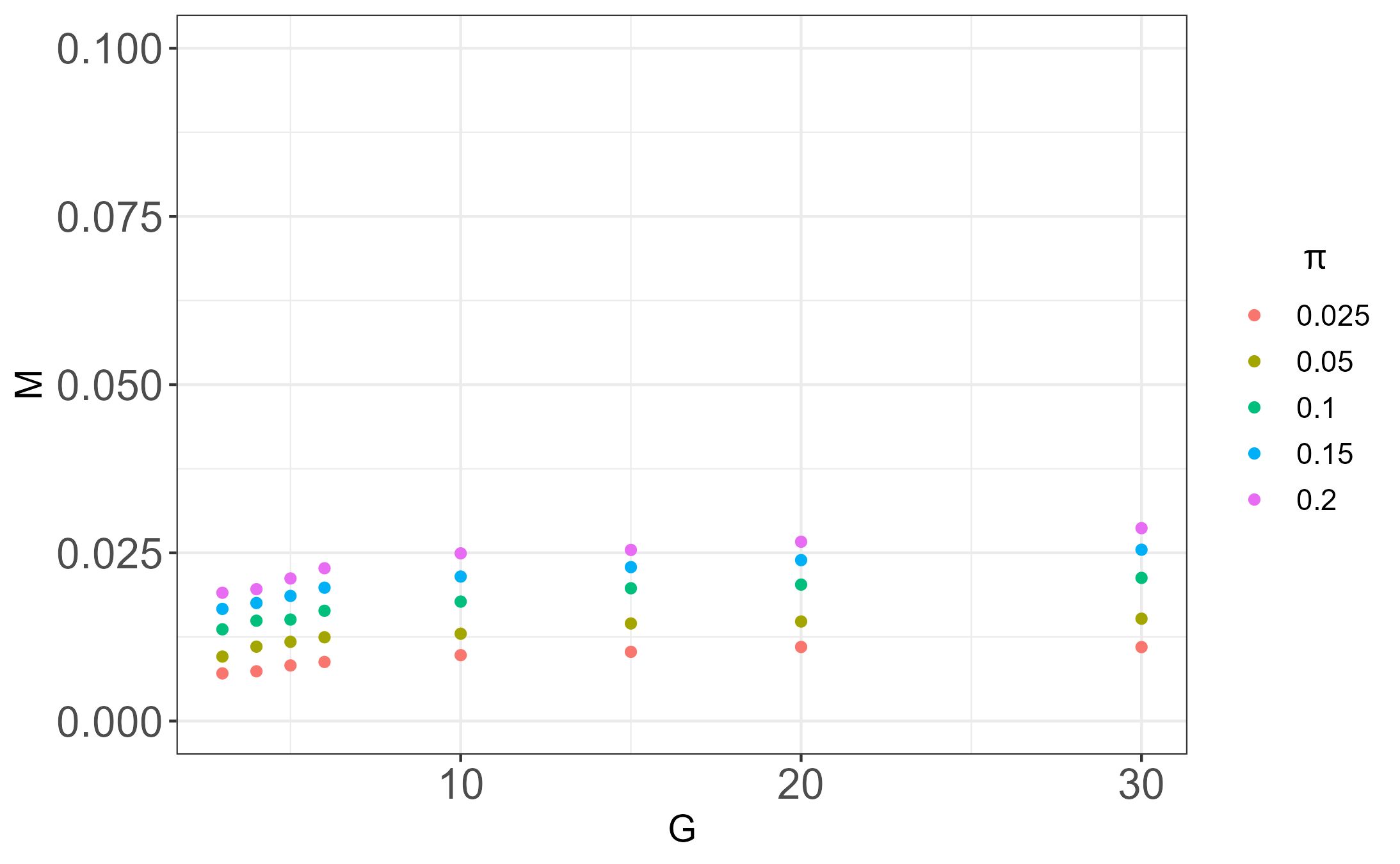}
         \hspace*{.4in}{\small (h) n=800, m=3, $\rho$=0.05}
    \label{fig:S3_Corr0.05_800}
    \end{minipage}
\begin{minipage}[htbp]{0.3\textwidth}
    \includegraphics[width=\textwidth]{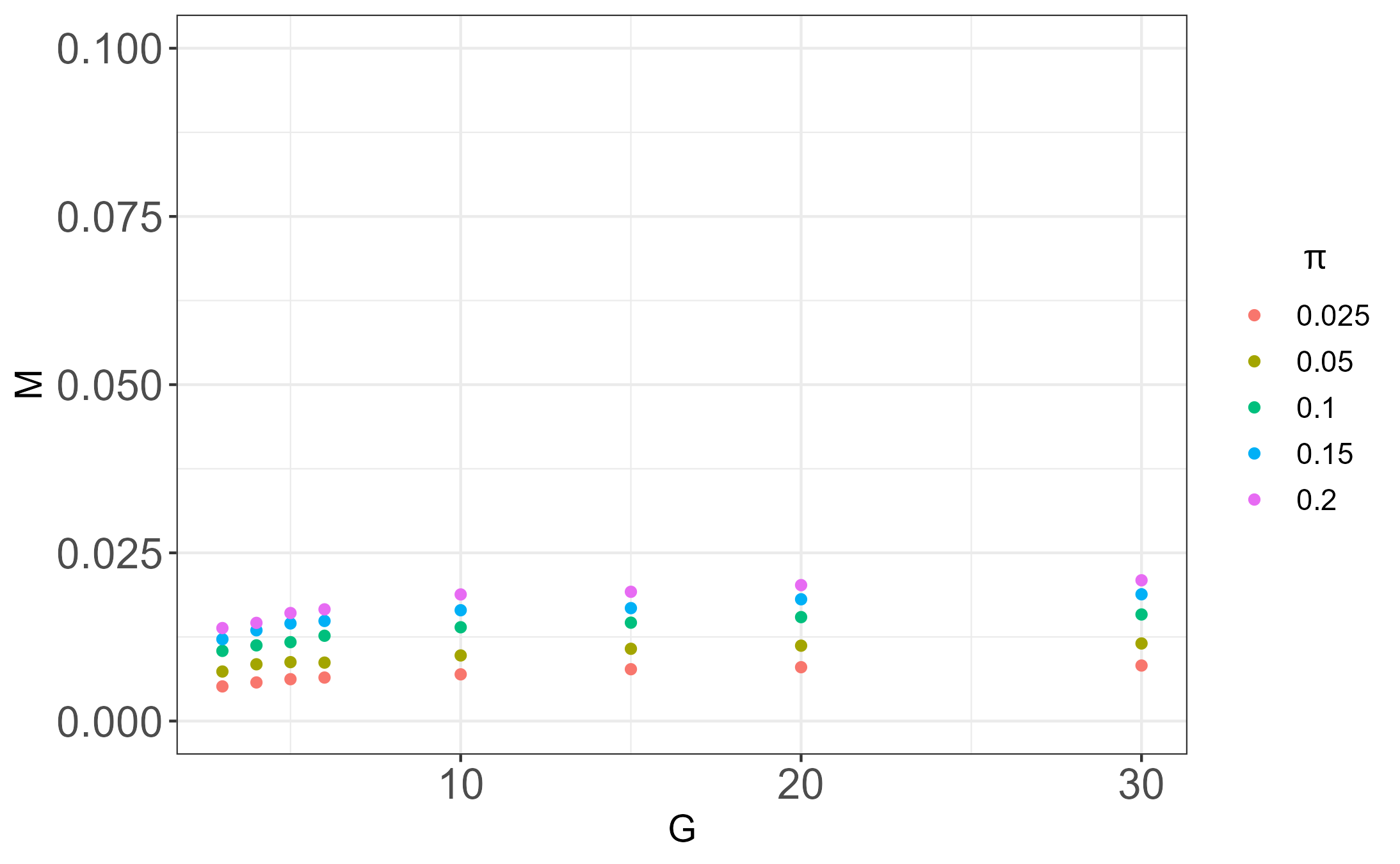}
         \hspace*{.4in}{\small (i) n=800, m=6, $\rho$=0.05}
    \label{fig:S6_Corr0.05_800}
\end{minipage}
\caption{ Results of simulation study for margin of error (M) as a function of number of individuals (n), number of attempts (m), correlation between attempts ($\rho$), and FNMR ($\pi$).  Subfigures are organized by columns where $m$ increases from left to right and by rows where $n$ increases from top to bottom.  Each figure plots $M$ versus $G$ for fixed $\rho$=0.05 and with different values for $\pi$ denoted by color.}
\label{broadAudience1}
\end{figure*}

\begin{figure*}[!t]
\centering

\textbf{Simulation Study Results for Margin of Error (M) versus Number of Groups (G)}
\par
\begin{minipage}[htbp]{0.35\textwidth}
    \includegraphics[width=\textwidth]{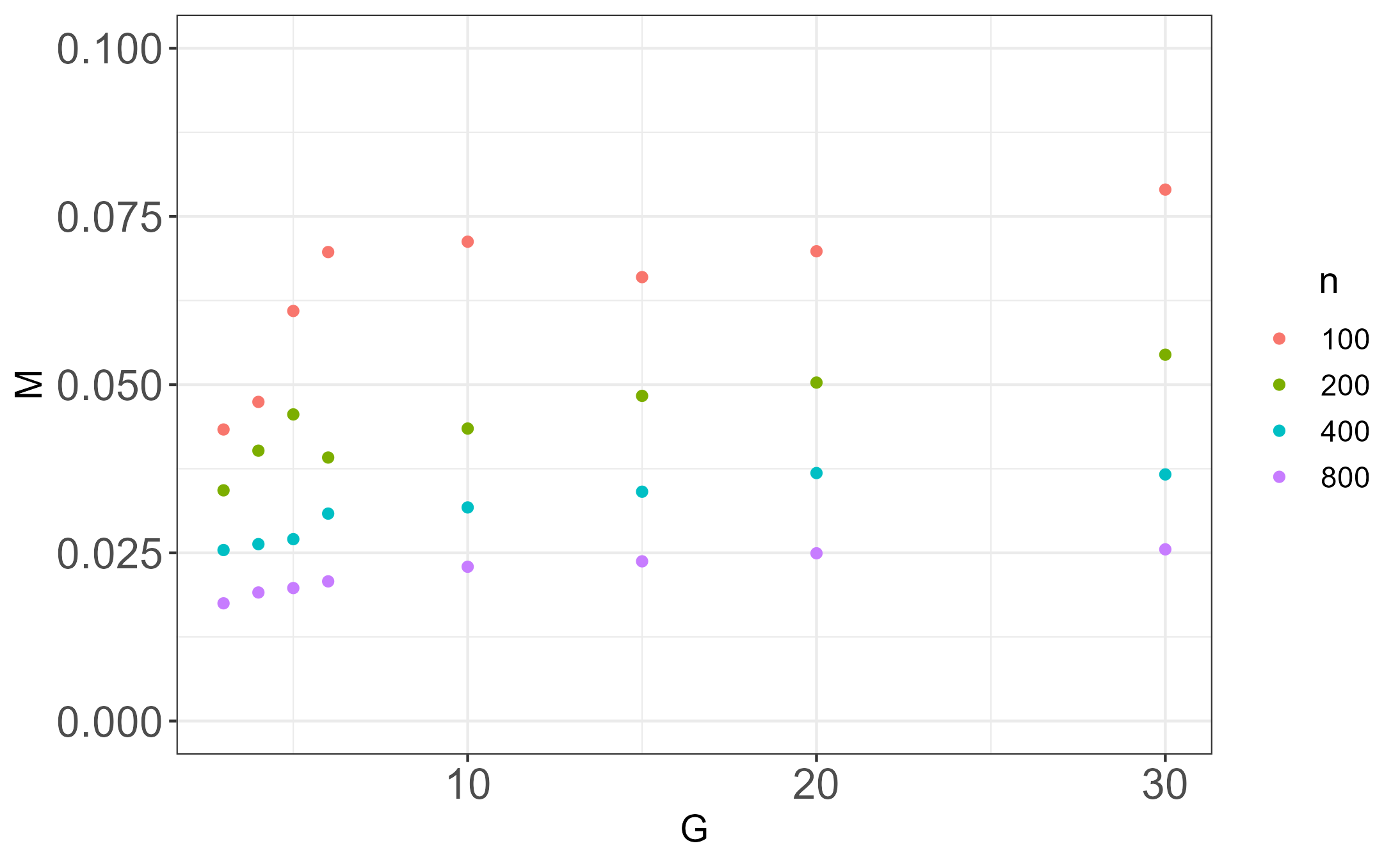}
   \hspace*{0.5in}{\small (a) $\pi$=0.1, m=2, $\rho$=0.05}
    \label{fig:m2_pi0.1_corr0.15}
    \end{minipage}
 \begin{minipage}[htbp]{0.35\textwidth}
    \includegraphics[width=\textwidth]{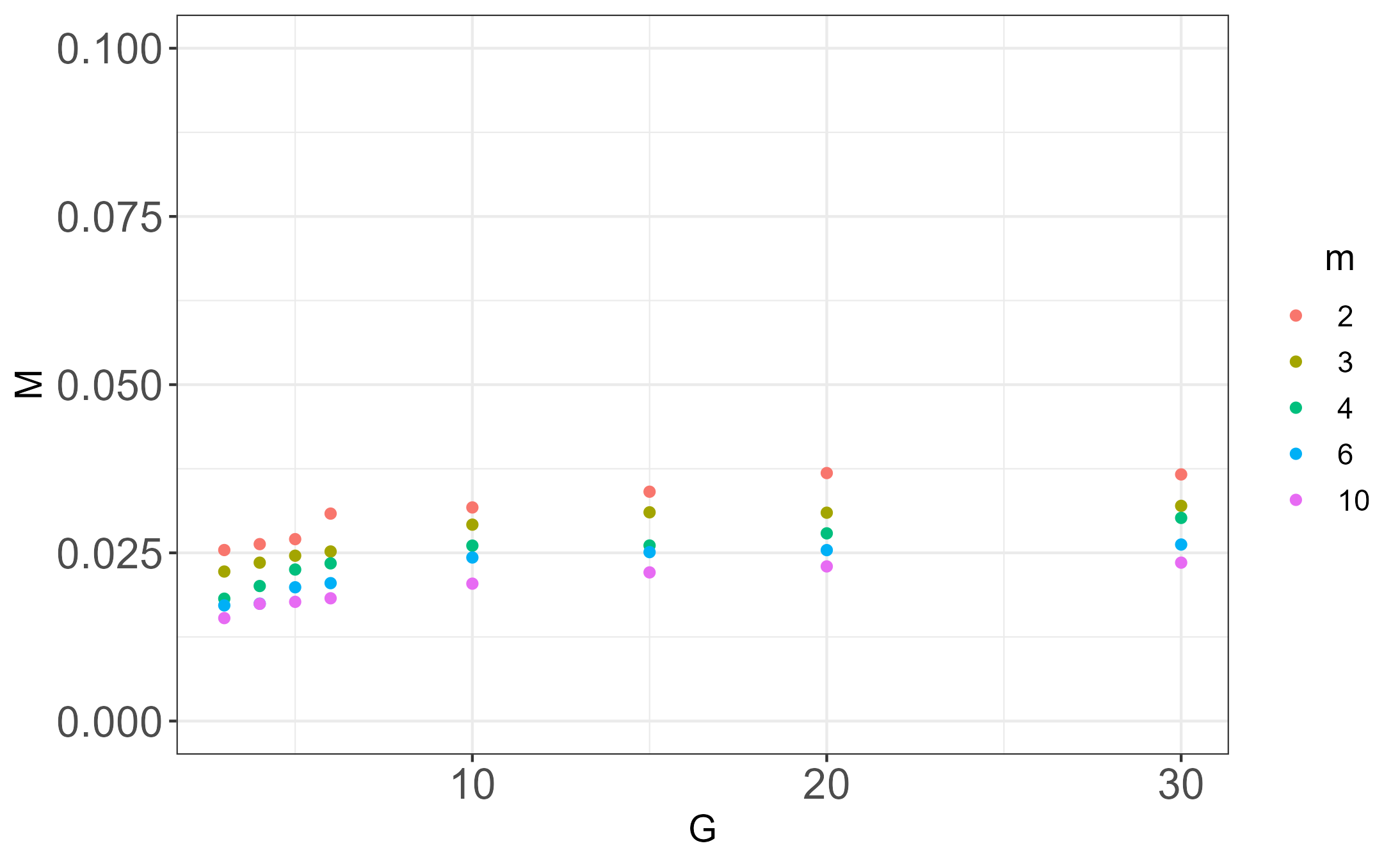}
    \hspace*{0.5in}{\small (b) n=400, $\pi$=0.1, $\rho$=0.05}
    \label{fig:Sub400_pi0.1_corr0.15}
    \end{minipage}
\hfill

\vspace{0.2in}
\begin{minipage}[htbp]{0.35\textwidth}
    \includegraphics[width=\textwidth]{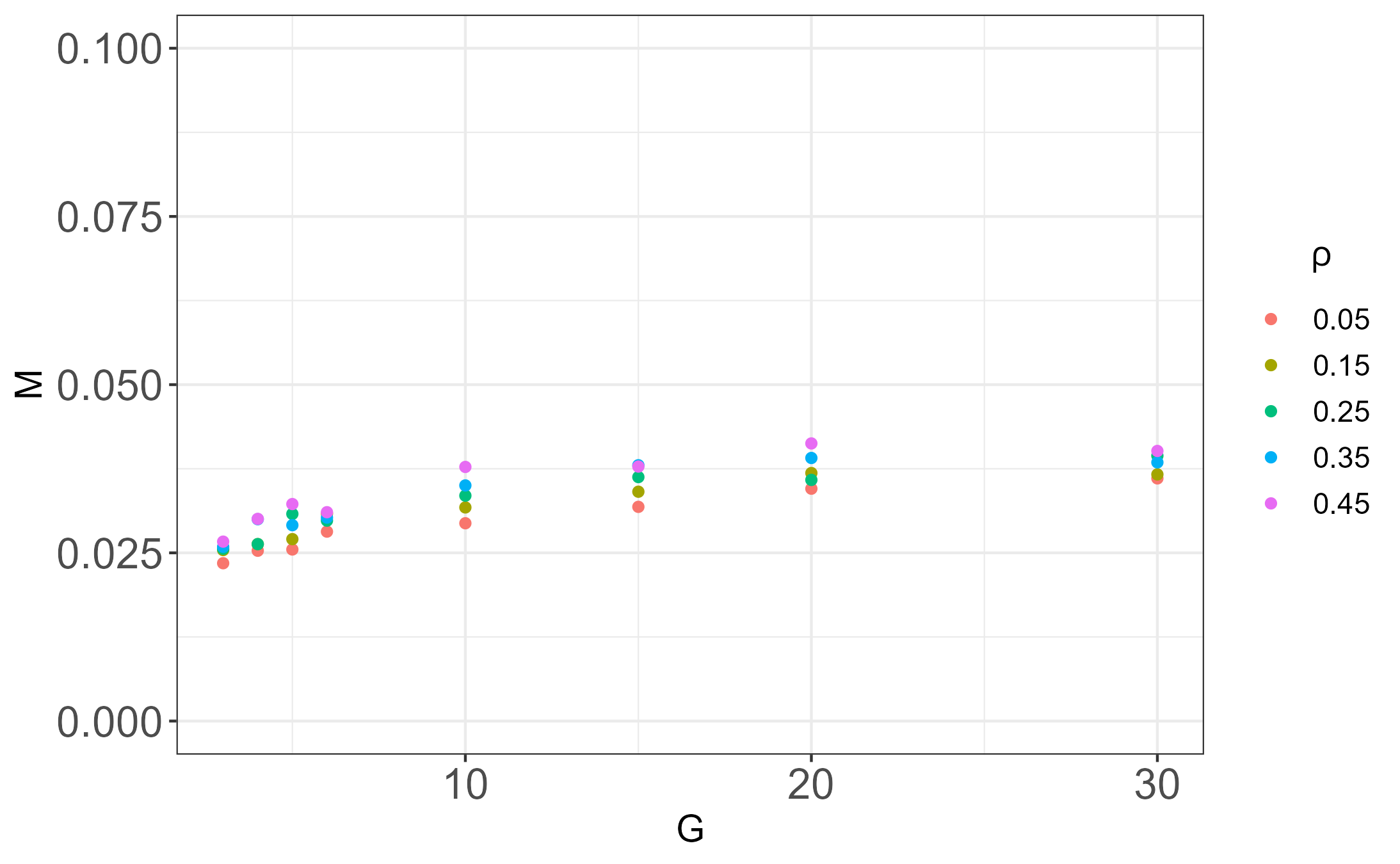}
   \hspace*{0.5in}{\small (c) n=400, m=2, $\pi$=0.1}
    \label{fig:m2_sub400_pi0.1}
\end{minipage}
\begin{minipage}[htbp]{0.35\textwidth}
    \includegraphics[width=\textwidth]{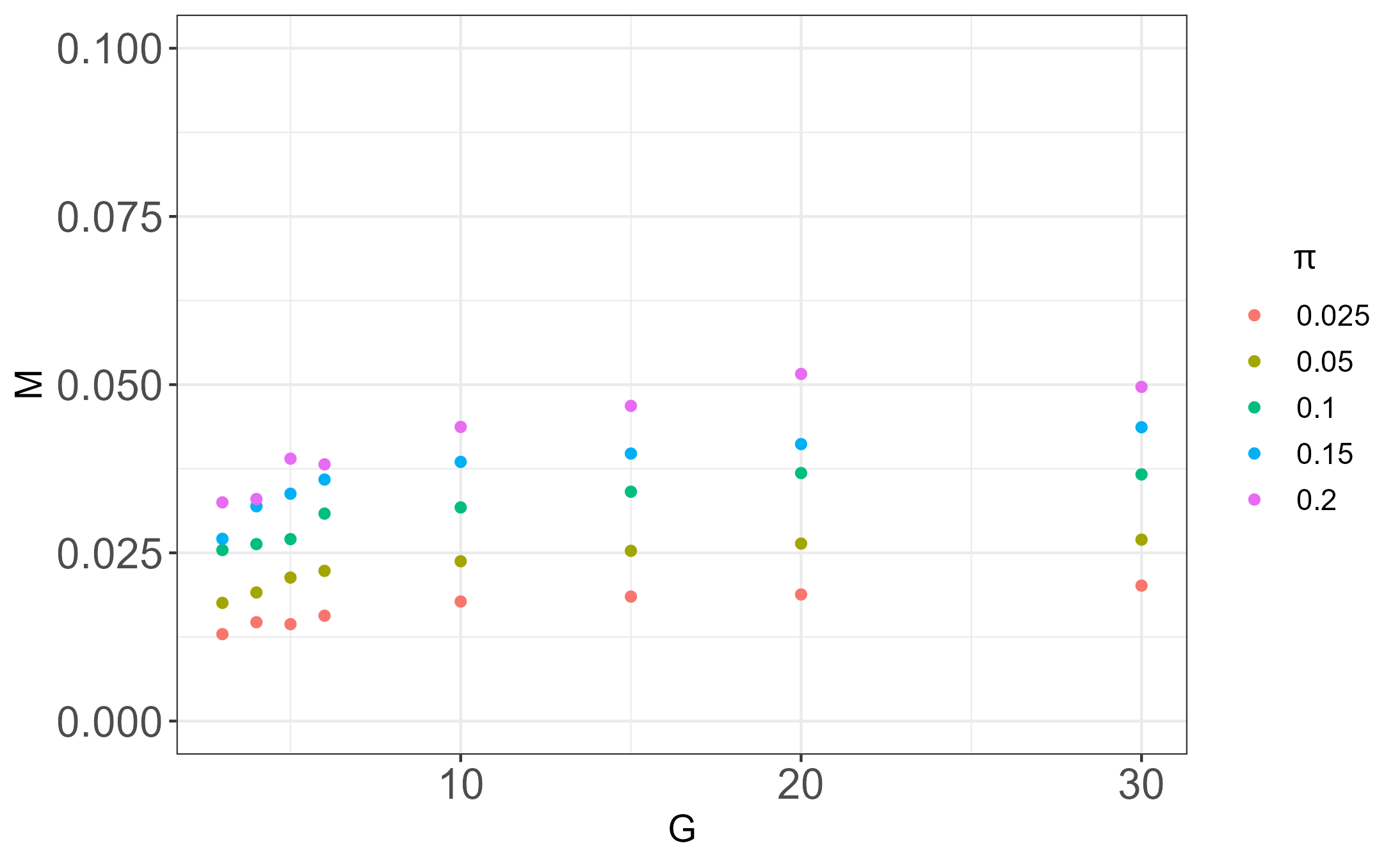}
    \hspace*{0.5in}{\small (d) n=400, m=2, $\rho$=0.05}
    \label{fig:m2_sub400_corr0.15}
\end{minipage}
\caption{Within each graph we vary different parameters from our simulation while fixing the others: each color is a varying n (subfigure a); varying m (b); varying $\rho$ (c); and varying $\pi$ (d) }
\label{broadAudience2}

\end{figure*}

In order to generate a single margin of error for all $G$ groups, we bootstrap the differences of each FNMR from the overall FNMR, then use the distribution of the maximal absolute differences to obtain $M$.  This approach is the following:
\begin{enumerate}
    \item Calculate the estimated overall FNMR, $\hat{\pi}$ and the estimated FNMR for 
    each group, $\hat{\pi}_g$, for $g=1,\ldots,G$.
    \item Sample with replacement the individuals in each group following Step 2) of the bootstrap approach above and calculate $\hat{\pi}_g^b$, the scaled bootstrapped estimated FNMM for group $g$.  
    \item Calculate and store $\phi = \max_g \vert \hat{\pi}_g^b - \hat{\pi} \vert$ using the notation from 
    Equation  \ref{eq:bootstrapdiff} of the bootstrap approach.
    \item Repeat the previous two steps $K$ times where $K$ is large, say more than $500$.
    \item  Determine $M$ by finding the $1-\alpha/2^{th}$ percentile from the distribution of $\phi$. 
\end{enumerate}

The maximal differences, the $\phi$'s, are calculated from each group FNMR which are scaled to (subtracted from) 
the overall mean,  so the distribution for $\phi$ that assumes variation if all of the FNMR's are equal. 
From this approach a range,$(\hat{\pi}-M, \hat{\pi}+M)$, of acceptable variation from the overall estimated FNMR, 
is generated.  The probability that a sample group FNMR 
would be outside of this interval by chance is $\alpha \times 100\%$ if all the groups are equal.  
To get a $100(1-\alpha)\%$ interval, use $\alpha = 0.05$. 
Thus, if $M=0.03$ is the $95^{th}$ percentile of the distribution of $\phi$'s 
and $\hat{\pi}=0.10$, the probability of a group $g$ having an FNMR be within $3\%$ of the overall FNMR is at least 90\%.  To illustrate this visually, consider Figure \ref{fig:illustrateM} where the top subfigure has no group FNMR's outside our interval while the bottom subfigure has three groups outside the generated bounds of $0.07$ and $0.13$. 
Thus, the practical use of this methodology is to produce an easily comprehensible range of values that would
not be different from  the overall FNMR and, likewise, yield a clearly delineated way to identify those groups with FNMR's that are statistical different from the overall mean.

\textit{Simulation Study}\\
To better understand how $M$ depends upon the parameters of our model, we performed a simulation study.  
Given $G$ groups, $n$ subjects/group, $m$ attempts per subject, $\pi$ as the FNMR rate, and $\rho$ as the correlation within subjects, we randomly generated false non match rates, $\hat{\pi}_g$, for each 
group $1000$ times and calculated $M$ as above.  We ran all combinations of the following values for each parameter: $\pi = 0.025, 0.05, 0.10, 0.15, 0.20$, $\rho = 0.05, 0.15, 0.25, 0.35, 0.45$, $n=100, 200, 400, 800$, \linebreak[5] $m = 2, 3, 4, 6, 10$ and $G = 3, 4, 5, 6, 10, 15, 20, 30$.  Our values for $\rho$ were selected to cover the values for estimated intra-individual correlations found in \cite{schuckers10}.  We fixed $\alpha$ at 0.05 for these simulations.  Figures \ref{broadAudience1}  and \ref{broadAudience2} show summaries of the results of these simulations with $M$ rounded to three decimal places with $\alpha=0.05$ in all cases.  Figure \ref{broadAudience1} 
shows simulation results for various values of $n$, the subfigure rows, $m$, the subfigure columns and $\pi$, colors within each subfigure while the intra-individual correlation $\rho$ was fixed at $0.05$ for these graphics. Within each subfigure, we have plotted $M$ versus $G$ and denoted different values of $\pi$ by different colors. 
From each subfigure, we can see that $M$ grows as $G$ increases though the amount of increase in $M$ slows as $G$ gets larger than $10$.  Moving down subfigure rows, ie.~ as $n$ increases we see that $M$ decreases.  Similarly, going from left to right across subfigure columns, i.e.~ as $m$ increases we see decreases in $M$.  Within each subfigure we can see that $M$ becomes smaller as $\pi$ decreases.  
In Figure \ref{broadAudience2}, we have plotted $M$ versus $G$ and varied a single parameter (denoted by different colors) in each subfigure at the values given above while fixing the other parameters at $n=400$, $\pi=0.10$, $\rho=0.05$ and $m=2$.  
From these values, we can see the impact of $n$, the number of individuals per group has the largest impact on $M$, followed by $\pi$, the overall FNMR, then $m$, the number of attempts per individual, and $\rho$ the intra-individual decision correlation.  Only $\rho$ is negatively associated with the size of $M$.  The impacts of $m$ and $\rho$ are tied together because of the nature of FNMR data.  While not shown in either of these figures, our simulations show that there is a positive, though not linear, relationship between $\rho$ and $M$.

\section{Discussion}

Equity and fairness in biometrics are important issues.  The declaration of differences between demographic groups is a consequential one.  Such conclusions about differences between groups need to be statistically 
sound and empirically based.  In this paper, we have proposed two approaches for 
testing for statistically detectable differences in FNMR's across $G$ groups.  
Our first approach uses the F-statistic as a metric and builds a reference distribution for that statistic via bootstrapping.  As mentioned above, this methodology, while valid and appropriate, is not easy to explain.  Our second approach attempts to remedy this drawback.  The second approach is to bootstrap  maximal differences among the FNMR's in the $G$ groups assuming a known equal FNMR across all groups, then generates a margin of error, $M$, to be added and subtracted to the overall FNMR for delineating which groups or subgroups are statistically different from the overall mean.  The latter approach has an advantage of being simpler and similar to other colloquial margins of error.  From our simulation study of this simpler approach, we have confirmed that the number of groups, $G$, and the number of individuals tested, $N$, substantially impact the margin of error.  Likewise though to a lesser extent the intra-individual correlation, $\rho$, and the number of attempts per individual, $m$, impact the size of $M$.
Our simplified approach uses the maximal absolute difference from the overall FNMR across the $G$ groups.   Using this distribution we generate an interval that is the overall FNMR plus and minus a margin of error $M$ where
$M$ is based upon the distribution of the maximal absolute difference.  Both of these methods, because they rely solely on thresholded decision data are applicable for testing commercial systems.

Both of our approaches in this paper have considered differences from the overall FNMR. but reasonable alternatives such as $\max_g (\frac{\hat{\pi}_g}{\hat{\pi}}, \frac{\hat{\pi}}{\hat{\pi}_g})$ might be of interest.  The importance of being able to generate a reference distribution to allow for an appropriate comparison to the observed statistics is critical to any statistical evaluation regardless of the functional form of the variation.  

This paper has looked at false non-match rate but similar methods and approaches exist for other common measures of bioauthentication performance including failure to enrol rates, failure to acquire rates and false match rates.  See \cite{schuckers10} for approaches for testing and comparing differences among multiple groups for these metrics.

\section*{Acknowledgment}
This work was supported by grants from the US National Science Foundation CNS-1650503 and CNS-1919554, and the Center for Identification Technology Research (CITeR).




\bibliographystyle{IEEEtran}
\bibliography{IEEEabrv.bib, IEEEexample.bib, goatssheep.bib}
%

\end{document}